\documentclass[atoms,article,accept,moreauthors]{Definitions/mdpi}
\usepackage{amsmath}
\usepackage{amsfonts}
\usepackage{amssymb,bm}
\usepackage{siunitx}
\usepackage{color}

\firstpage{1}
\pubvolume{9}
\issuenum{1}
\articlenumber{4}
\pubyear{2021}
\copyrightyear{2021}
\history{Received: 11 December 2020; Accepted: 11 January 2021; Published: 14 January 2021}

\Title{Casimir-Polder Interaction of an Atom with a Cavity Wall Made of
        Phase-Change Material out of Thermal Equilibrium }

\Author{Galina L. Klimchitskaya $^{1,2}$ and Vladimir M. Mostepanenko $^{1,2,3}$
}

\address{%
$^{1}$ \quad Central Astronomical Observatory at Pulkovo of the Russian Academy of Sciences, Saint Petersburg, 196140, Russia\\
$^{2}$ \quad Institute of Physics, Nanotechnology and
Telecommunications, Peter the Great Saint Petersburg
Polytechnic University, Saint Petersburg, 195251, Russia\\
$^{3}$ \quad Kazan Federal University, Kazan, 420008, Russia}

\corres{Correspondence: vmostepa@gmail.com}

\abstract{We consider the out-of-thermal-equilibrium Casimir-Polder
interaction between atoms of He$^*$, Na, Cs, and Rb and a cavity wall
made of sapphire coated with a vanadium dioxide film which undergoes the
dielectric-to-metal phase transition with increasing wall temperature.
Numerical computations of the Casimir-Polder force and its gradient
as the functions of atom-wall separation and wall temperature are made when
the latter exceeds the temperature of the environment. The
obtained results are compared with those in experiment on measuring the
gradient of the Casimir-Polder force between $^{87}$Rb atoms and a silica
glass wall out of thermal equilibrium. It is shown that the use of phase-change
wall material increases significantly the force magnitude and
especially the force gradient, as opposed to the case of dielectric wall.
}

\keyword{Atom-wall interaction; Atomic polarizability;
Nonequilibrium Casimir-Polder force; Phase-change material;
Dielectric-to-metal phase transition}

\begin{document}

\section{Introduction}

The interaction of atoms with material surfaces is a long-explored
area. At atom-surface separations exceeding one or two nanometers it is
determined to a large extent by the zero-point and thermal fluctuations
of the electromagnetic field. This is a particular case of the van der
Waals forces which act between two atoms, an atom and a surface or
between two surfaces and originate from fluctuating electric dipole
moments \cite{1}. The first theory of atom-surface interaction was
created by London on the basis of nonrelativistic quantum mechanics
\cite{2}. It was generalized taking into account an existence of
the zero-point fluctuations and retardation of the electromagnetic
interaction by Casimir and Polder who studied a polarizable atom in
the vicinity of an ideal metal wall \cite{3}. Presently the atom-wall
force is often referred to by their names.

A more general theory taking into account the material properties
of the wall at nonzero temperature in thermal equilibrium with the
environment was developed in \cite{4} on the basis of the Lifshitz
theory \cite{5,6}. In the framework of this theory, the free energy
of the Casimir-Polder interaction and respective force were expressed
via the frequency-dependent atomic polarizability and dielectric
permittivity of the wall material. In succeeding years the fluctuation-induced
forces between atoms, molecules and material surfaces found
numerous applications in atomic physics, condensed matter physics, as
well as in biology and chemistry \cite{1,7}. The advent of magnetic
traps and ultracold cooling has opened up new opportunities for
detailed study of the Casimir-Polder force in high-precision
experiments.

Thus, the phenomenon of quantum reflection, i.e.,
reflection of ultracold atoms under an action of the attractive
atom-surface force was demonstrated for both liquid and solid
surfaces \cite{8,9,10,11,12,13,14,15,16,17}. Another phenomenon is
the Bose-Einstein condensation in dilute gases cooled to very low
temperatures and confined in the magnetic trap near a material
surface \cite{18,19,20,21,22}. Although the Casimir-Polder force
between a unit atom and a surface is extremely small and is not
directly measurable, it should be accounted for in precise
experiments with atomic clouds involving an abundance of atoms.
This stimulated applications of the Lifshitz theory to calculation
of the Casimir-Polder force for different atoms and cavity walls
taking into account the atomic dynamic polarizabilities and
dielectric properties of wall materials \cite{23,24,25,26,27,28,
29,30,31,32,33,34,35,36,37,38,39,40,41,42,43,44,45}.

During the last few years special attention was paid to the
nonequilibrium Casimir-Polder forces which arise in situations
when the surface (cavity wall) and the environment are kept at
different temperatures. It is usually assumed that the environment
temperature remains unchanged whereas the cavity wall is heated to
some higher temperature. The Casimir-Polder force under these
conditions was discussed in \cite{46}, whereas the generalization
of the Lifshitz theory to nonequilibrium situations, including the
case of atom-wall interaction, was performed in \cite{47,48}. This
theory was used for interpretation of the measurement data of the
experiment \cite{49} where the thermal Casimir-Polder force between
the ground state $^{87}$Rb atoms, belonging to the Bose-Einstein
condensate, and a silica glass wall heated up to 605~K has been
measured for the first time. The measure of agreement between the
experimental data of \cite{49} and theoretical predictions was used
for constraining the Yukawa-type corrections to Newtonian gravity
\cite{50} and the coupling constants of axions to nucleons \cite{51}.

In this article, using the formalism of \cite{47,48}, we consider
another physical situation where the nonequilibrium effects are
gaining in importance, i.e., the Casimir-Polder force between
different atoms and subjected to heating cavity wall which is made
of the phase-change material. In fact the equilibrium Casimir forces
acting between a Au-coated sphere and either a VO$_2$ film deposited
on a sapphire substrate \cite{52} or a AgInSbTe film  on a Si
substrate \cite{53} have already been considered in the literature.
It has been known that VO$_2$ undergoes the phase transition from
the dielectric to metallic phase with increasing temperature above
$T_c = 341~$K. As to an amorphous AgInSbTe film, it undergoes the
phase transition to the crystalline phase as a result of annealing.
It was shown that the equilibrium Casimir force between a Au sphere
and a VO$_2$ film experiences an abrupt jump resulting from the phase
transition \cite{52}. The same holds for the Casimir force between
a sphere and an AgInSbTe film in different phase states \cite{53}.
Thus, it would be interesting to consider the combined effect of the
nonequilibrium and phase-change conditions on the Casimir-Polder
force for the wall material which undergoes the phase transition with
increasing temperature.

Here, we consider the atoms of metastable helium He$^*$, Na, Rb, and
Cs interacting with a VO$_2$ film deposited on a sapphire wall for the
atom-film separations varying from 5 to 10 $\mu$m much as in
the experiment \cite{49} for $^{87}$Rb atoms and a SiO$_2$ wall. We
calculate the nonequilibrium Casimir-Polder forces for all these
atoms at different wall temperatures, both below and above the
critical temperature $T_c$ at which the VO$_2$ film transforms from
the dielectric to metallic phase, leaving the environment temperature
constant. It is shown that the magnitude of the Casimir-Polder force
decreases monotonously with increasing atom-wall separation, but
takes much larger value at each separation if the wall temperature,
although slightly, exceeds the critical temperature. When the atom-
wall separation is fixed, the force magnitude increases monotonously
with increasing wall temperature in the region $T < T_c$, experiences a
jump at $T = T_c$, and increases further at $T > T_c$.

By way of example, for a Rb atom 5 $\mu$m apart from the wall made
of the phase change-material the magnitudes of equilibrium Casimir-Polder
force at 300~K, 340~K (i.e., before the phase transition), and
342~K (i.e., after the phase transition) are equal to 1.92, 2.17, and
2.68 ($10^{-13}$fN), respectively. If, however, the environment
temperature remains equal to 300~K and only the wall is heated, one
obtains the force magnitudes of 1.92, 5.05, and 8.29 ($10^{-13}~$fN)
at the same respective wall temperatures. We have also compared the
nonequilibrium Casimir-Polder forces and their gradients  between a Rb atom and
either a VO$_2$ film on a sapphire wall, which undergoes the phase transition,
or a dielectric SiO$_2$ wall employed in the experiment \cite{49}.
Using the experimental parameters, it is shown, for instance, that the
nonequilibrium Casimir-Polder force between a Rb atom 7 $\mu$m apart
from the wall made of the phase-change material reaches at $T = 415~$K
wall temperature the same magnitude as is reached at $T = 605~$K in
the case of a dielectric wall.

The structure of this article is the following. In Section~2, we
briefly outline the used formalism. Section~3 contains the
calculation results for the nonequilibrium Casimir-Polder force
for different atoms and phase-change wall material as the functions
of atom-wall separation and wall temperature. In Section~4, the
nonequilibrium Casimir-Polder force and its gradient for the phase-change
wall material are compared with those for a dielectric wall used in the
experimental configuration \cite{49}. Section~5 contains a
discussion of the obtained results. In Section~6 the reader will
find our conclusions.


\section{Nonequilibrium Casimir-Polder Force in the Micrometer
Separation Range}

Here, we summarize the main expressions, allowing calculation of the
nonequilibrium force between a ground-state atom and a cavity wall spaced
at separations of a few micrometers, obtained in \cite{47,48}.
In doing so atom is described by its static polarizability $\alpha(0)$
and wall material  by its dielectric properties at vanishing frequency.
In the case of phase-change wall material, these properties are very
specific and endow the nonequilibrium Casimir-Polder force with new useful
applications.

Let the temperature of the environment be $T_E=300~$K and the wall temperature
$T_W\geqslant T_E$. The atom-wall separation is denoted by $a$.
Then the nonequilibrium Casimir-Polder force can be presented in the form
\cite{47,48}
\begin{equation}
F(a,T_W,T_E)=F_{\rm eq}(a,T_E)+F_{\rm neq}(a,T_W) -F_{\rm neq}(a,T_E).
\label{eq1}
\end{equation}
\noindent
Here, $F_{\rm eq}$ is the well known equilibrium Casimir-Polder force given
by the Lifshitz formula and $ F_{\rm neq}$ is the proper nonequilibrium
contribution. As is seen from (\ref{eq1}), in the state of thermal equilibrium
$T_W=T_E$, the nonequilibrium terms cancel each other, and the Casimir-Polder
force reduces to the equilibrium contribution alone.

The explicit expressions for both terms on the right-hand side of (\ref{eq1})
at all separations can be found in \cite{47,48} (see also the monograph \cite{54}).
Keeping in mind the values of the experimental parameters \cite{49}, here we
deal with atom-wall separations in the region from 5 to 10~$\mu$m.
In this region one can obtain simple asymptotic expressions for both  $F_{\rm eq}$
and $F_{\rm neq}$. Thus, for an ideal dielectric wall possessing finite dielectric
permittivity at zero frequency  $\varepsilon(0)<\infty$ one has \cite{54,55}
\begin{equation}
F_{\rm eq}(a,T)=-\frac{3k_B T}{4a^4}\,\alpha(0)\,
\frac{\varepsilon(0)-1}{\varepsilon(0)+1},
\label{eq2}
\end{equation}
\noindent
where $k_B$ is the Boltzmann constant.

It is common knowledge, however, that at any nonzero temperature real dielectric
materials possess some small conductivity (the so-called dc conductivity) which vanishes
with $T$ exponentially fast \cite{56,57}. As a result, for real dielectric
materials one obtains $\varepsilon(\omega)\to\infty$ when $\omega$ goes to zero
and instead of (\ref{eq2}) one has
\begin{equation}
F_{\rm eq}(a,T)=-\frac{3k_B T}{4a^4}\,\alpha(0).
\label{eq3}
\end{equation}
\noindent
The same result was derived for the equilibrium Casimir-Polder force between
an atom and a metallic wall at separations exceeding a few micrometers.
Note that (\ref{eq3}) is given by the zero-frequency term of the Lifshitz
formula. In so doing, the omitted sum over all nonzero Matsubara frequencies,
which describes contribution of the zero-point oscillations to the force and
some extra thermal contribution, is equal to 6.8\% of (\ref{eq3}) at
$a=5~\mu$m, $T=300~$K and quickly decreases with increasing separation
and/or temperature.

Equations (\ref{eq2}) and (\ref{eq3}) invite some further information.
The point is that measurements of the equilibrium Casimir force between a
Au-coated sphere and a dielectric plate at separations of a few hundred
nanometers were found in disagreement with theoretical predictions of the
Lifshitz theory if the dc conductivity of plate material is taken into
account in computations \cite{58,59,60,61}. An agreement between the theory and
the measurement data is reached only if the dc conductivity is disregarded
\cite{58,59,60,61}. The measured equilibrium Casimir-Polder force is also in
good agreement with theoretical predictions if the dc conductivity of
SiO$_2$ wall is omitted in computations \cite{49}, but excludes these predictions
found with taken into account dc conductivity \cite{62}. What is even more
surprising, the Casimir-Polder entropy corresponding to the equilibrium
Casimir-Polder force found in the framework of the Lifshitz theory violates
the Nernst heat theorem if the dc conductivity of the dielectric wall is
included in calculation \cite{63,64,65,66,67}. This conundrum is not yet resolved.
Because of this in computations below we consider both cases of taken into
account and disregarded dc conductivity of the cavity wall at temperatures
below $T_c$ when VO$_2$ is in the dielectric phase.

A similar problem was revealed for the equilibrium Casimir force measured between
metallic sphere and metallic plate. It was found that theoretical
predictions of the Lifshitz theory are excluded by the measurement data if the
relaxation properties of conduction electrons are taken into account in
calculations (see, e.g., \cite{54,55,68,69} for a review and \cite{70} for an
attempts to solve this puzzle). For the interaction of nonmagnetic atoms with
metallic wall, however, the equilibrium Casimir-Polder force given by the Lifshitz
formula does not depend on the relaxation properties of conduction electrons and,
thus, the problem becomes immaterial.

Now we consider the nonequilibrium contributions on the right-hand side of (\ref{eq1}).
At separations of a few micrometers between an atom and a dielectric wall, the
asymptotic expression for $F_{\rm neq}$ was derived under a condition that the
thermal frequency $\omega_T=k_B T/\hbar$ is much less than the characteristic
frequency of the wall material \cite{47,71} (see also review \cite{71a})
\begin{equation}
F_{\rm neq}(a,T)=-\frac{\pi\alpha(0)(k_B T)^2}{6c\hbar a^3}\,
\frac{\varepsilon(0)+1}{\sqrt{\varepsilon(0)-1}}.
\label{eq4}
\end{equation}

This result was obtained with disregarded dc conductivity of the dielectric wall
$\sigma_0(T)$ connected with the dielectric permittivity according to
\begin{equation}
\varepsilon(\omega)=\varepsilon(0)+\frac{4\pi {\rm i}\sigma_0(T)}{\omega}.
\label{eq5}
\end{equation}
\noindent
We emphasize, however, that for the nonequilibrium contribution to the Casimir-Polder
force the inclusion in the calculation of the dc conductivity
$\sigma_0(T)\ll\omega_T$ leads to practically the same values of  $F_{\rm neq}$
as are given in (\ref{eq4}) \cite{62}. This is distinct from the equilibrium contribution
to the Casimir-Polder force between an atom and a dielectric wall.

For an atom interacting with a metallic wall possessing the conductivity $\sigma_m(T)$ under
a condition $\omega_T\ll\sigma_0(T)$ it holds \cite{47}
\begin{equation}
F_{\rm neq}(a,T)=-\frac{\alpha(0)\zeta(3/2)\sqrt{\sigma_m(T)\,}
(k_B T)^{3/2}}{c\sqrt{2\hbar}\, a^3},
\label{eq6}
\end{equation}
\noindent
where $\zeta(z)$ is the Riemann zeta function. It should be stressed that this
equation is derived taking into account the relaxation properties of conduction
electrons by means of the Drude model. If the relaxation properties of conduction
electrons in metals are disregarded, the theoretically meaningless result for
$F_{\rm neq}$ follows which is also in contradiction with the measurement data.
Because of this one can conclude that the puzzling problems discussed above refer to only
the Casimir and Casimir-Polder forces in the state of thermal equilibrium.
Note also that following \cite{47,71} we consider atoms in their ground states and
assume that they cannot absorb the thermal radiation. Possible effects caused by
the atomic absorption spectrum and radiation pressure are discussed in \cite{71a}.

In the next sections, the above results are applied to different atoms
interacting with a wall made of the phase-change material.

\section{The Casimir-Polder Force between Different Atoms and
{\boldmath VO$_2$}  Film on a Sapphire Wall}

Here, we consider the  atoms of metastable He$^{\ast}$ and ground state Na, Rb,
and Cs interacting with VO$_2$ film of thickness 100~nm deposited on bulk sapphire
wall. The atoms under consideration are characterized by their static atomic
polarizabilities with the following values:\hfill\\
$\alpha^{\rm He^{\ast}}(0)=4.678\times 10^{-29}~\mbox{m}^3$ \cite{72,73},
$\quad\alpha^{\rm Na}(0)=2.411\times 10^{-29}~\mbox{m}^3$ \cite{74},\hfill\\
$\alpha^{\rm Rb}(0)=4.73\times 10^{-29}~\mbox{m}^3$ \cite{75}, $\quad$and
$\quad\alpha^{\rm Cs}(0)=5.981\times 10^{-29}~\mbox{m}^3$ \cite{74,76}.

It is  known that both vanadium dioxide (VO$_2$) crystals and thin films undergo
the phase transition from dielectric monoclinic to a metallic tetragonal phase when
temperature increases up to $T_c=341~$K \cite{77}. The dielectric permittivity of
100~nm thick VO$_2$ film deposited on a sapphire wall was measured in the frequency
region from $3.8\times 10^{14}$ to $7.6\times 10^{15}~$rad/s and fitted to the
isotropic oscillator representation in \cite{78,79} both before and after the
phase transition.

According to the obtained results, the static dielectric permittivity of VO$_2$ film
in the dielectric phase on a sapphire wall is equal to $\varepsilon(0)=9.909$.
The characteristic absorption frequency of a VO$_2$ film on a sapphire wall before
the phase transition is of the order of $1.5\times 10^{15}~$rad/s, i.e., it is much
larger than the thermal frequency $\omega_T=3.9\times 10^{13}~$rad/s at room temperature
$T=300~$K and also at higher temperatures up to 600~K considered below. Because of this,
Equation~(\ref{eq4}) is applicable in this case. The dc conductivity in the dielectric
phase if of the order of $\sigma_0\sim 10^{11}~\mbox{s}^{-1}$. In the region around
room temperature it is almost temperature-independent but goes to zero exponentially
fast with vanishing $T$.  As was noted in Section~2, an account of this conductivity
does not make an impact on the result (\ref{eq4}).

In the metallic phase at $T=355~$K the conductivity value of the film was measured to be
$\sigma_m(T)=2.03\times 10^{15}~\mbox{s}^{-1}$ \cite{78}, i.e., by the four orders
of magnitude higher than the dc conductivity in the dielectric phase.
The conductivity $\sigma_m(T)$ is well described in the framework of the Drude model as
\begin{equation}
\sigma_m(T)=\frac{\omega_p^2}{4\pi\gamma(T)},
\label{eq7}
\end{equation}
\noindent
where $\omega_p=5.06\times 10^{15}~$rad/s is the plasma frequency and
$\gamma(T=355\,\mbox{K})=1.0\times 10^{15}~$rad/s is the relaxation parameter.
In the temperature region of our interest the linear dependence of the relaxation
parameter on $T$ is determined by the electron-phonon interaction \cite{80}
\begin{equation}
\gamma(T)=k T,\qquad k=\frac{1.0\times10^{15}}{355}\,
\frac{\mbox{rad}}{\mbox{s\,K}}.
\label{eq8}
\end{equation}
\noindent
It is seen that for VO$_2$ film in a metallic phase $\omega_T\ll\sigma_m(T)$ when
$T$ varies between 300~K and 600~K. Thus, the application condition of
Equation~(\ref{eq6}) is satisfied.

Now we compute the Casimir-Polder force between a Rb atom and a VO$_2$ film on
a sapphire wall as a function of separation between them assuming that the
temperature of the environment is $T_E=300~$K whereas the wall temperature $T_W$
is either equal  to $T_E$ (the situation of thermal equilibrium) or $T_W>T_E$
(the out-of-thermal-equilibrium situation), specifically, $T_W=340~$K, 345~K,
and 385~K. In doing so the first of these temperatures corresponds to a VO$_2$
film in the dielectric phase, whereas the next two --- in the metallic phase.

\begin{figure}[!ht]
\centering
\vspace*{-2.5cm}
\includegraphics[width=15 cm]{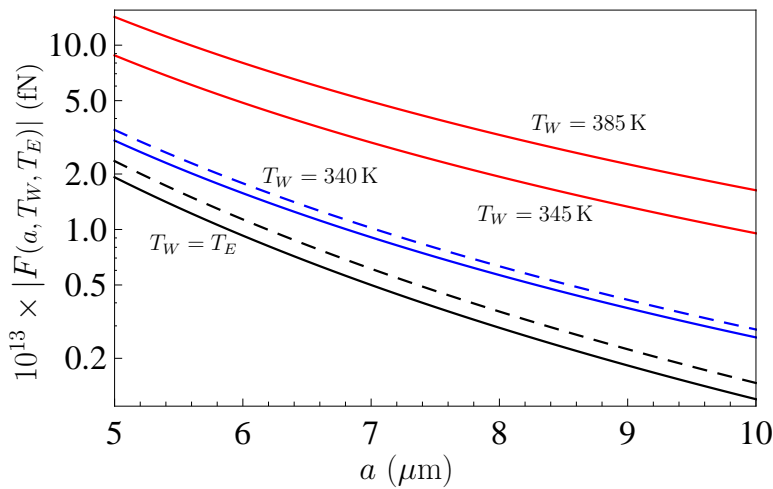}
\vspace*{-12cm}
\caption{The magnitudes of the Casimir-Polder force between Rb
atoms and VO$_2$ film on a sapphire wall are shown in the logarithmic
scale by the solid and dashed lines as the functions of separation at
the environment temperature $T_E = 300~$K and the following wall
temperatures $T_W$: $T_W = T_E$ (thermal equilibrium, the lower pair
of solid and dashed lines computed with disregarded and taken into
account dc conductivity, respectively), $T_W = 340~$K (out of thermal
equilibrium, the upper pair of solid and dashed lines computed with
disregarded and taken into account dc conductivity, respectively),
$T_W = 345~$K (out of thermal equilibrium, metallic phase, the lower
of two top solid lines), $T_W = 385~$K (out of thermal equilibrium,
metallic phase, the upper of two top solid lines).
\label{fig1}}
\end{figure}
The computational results for the force magnitude are shown in Figure~\ref{fig1}
in the logarithmic scale as functions of separation by the lower and upper pairs
of solid and dashed lines found at $T_W=T_E=300~$K and  $T_W=340~$K,
respectively, and by the next to them two solid lines computed at $T_W=345~$K and
385~K. At thermal equilibrium (the lower pair of solid and dashed lines)
the computations are made by Equations (\ref{eq2}) and (\ref{eq3}), respectively,
with disregarded and taken into account dc conductivity. The upper pair of solid
and dashed lines (out-of-thermal-equilibrium situation but VO$_2$ film is in the
dielectric phase) are computed by Equations (\ref{eq1}), (\ref{eq2}), (\ref{eq4})
and (\ref{eq1}), (\ref{eq3}), (\ref{eq4}), respectively,
with disregarded and taken into account dc conductivity in the equilibrium
contribution. The top two solid lines in Figure~\ref{fig1}
(out-of-thermal-equilibrium situation but VO$_2$ film is in the
metallic phase) are computed by Equations (\ref{eq1}), (\ref{eq3}), and
(\ref{eq6})--(\ref{eq8}) at respective temperatures. Note that (\ref{eq6}) and
(\ref{eq8}) take into account the dependence of the conductivity of metal on
temperature as a parameter as it should be done also for nonequilibrium
Casimir force between two parallel plates \cite{81}.

As is seen in Figure~\ref{fig1}, the magnitude of negative (attractive)
Casimir-Polder force decreases with increasing atom-wall separation and
increases with increasing temperature. If the dc conductivity is taken into
account in computations (the dashed lines in the lower and upper pairs of lines
related to a VO$_2$ film in the dielectric phase), the larger magnitudes of
the Casimir-Polder force are obtained. There is a jump in the force magnitude
in the temperature region from  $T_W=340~$K to 345~K which includes the
critical temperature $T_c=341~$K of the phase transition.

\begin{figure}[!t]
\centering
\vspace*{-2.5cm}
\includegraphics[width=15 cm]{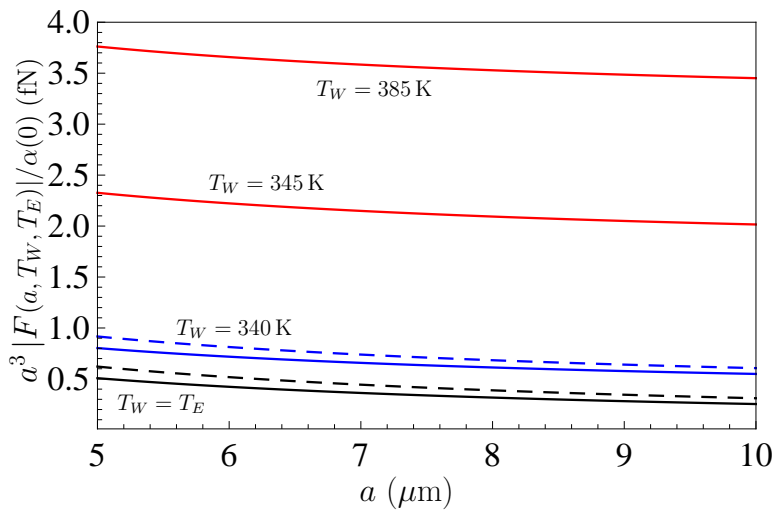}
\vspace*{-12cm}
\caption{The magnitudes of the Casimir-Polder force between Rb
atoms and VO$_2$ film on a sapphire wall times the factor
$a^3/\alpha^{Rb}(0)$ are shown as the functions of separation at
the environment temperature $T_E = 300~$K and the following wall
temperatures $T_W$: $T_W = T_E$ (thermal equilibrium, the lower pair
of solid and dashed lines computed with disregarded and taken into
account dc conductivity, respectively), $T_W = 340~$K (out of thermal
equilibrium, the upper pair of solid and dashed lines computed with
disregarded and taken into account dc conductivity, respectively),
$T_W = 345~$K (out of thermal equilibrium, metallic phase, the lower
of two top solid lines), $T_W = 385~$K (out of thermal equilibrium,
metallic phase, the upper of two top solid lines).
\label{fig2}}
\end{figure}

To gain a better insight, in Figure~\ref{fig2} we present the same computational
results for the magnitude of the Casimir-Polder force, as in Figure~\ref{fig1},
times the factor $a^3/\alpha^{\rm Rb}(0)$. This allows to plot the figure in
a homogeneous scale preserving the meaning and disposition of all lines.
What is more, Figure~\ref{fig2} does not depend on the type of atom and gives
the possibility to obtain the magnitudes of the nonequilibrium Casimir-Polder
force for any atom interacting with a VO$_2$ film deposited on a sapphire wall,
and, specifically, for the atoms of He$^{\ast}$, Na, and Cs. For this purpose,
one should multiply the data of Figure~\ref{fig2} by the factor
$\alpha(0)/a^3$ using the value of  $\alpha(0)$ for the desirable atom.

\begin{figure}[!b]
\centering
\vspace*{-2.5cm}
\includegraphics[width=15 cm]{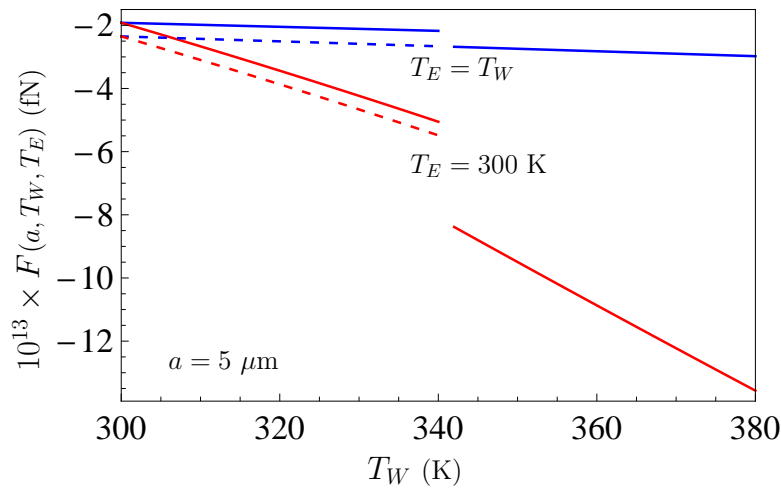}
\vspace*{-12cm}
\caption{The Casimir-Polder forces between Rb atoms and VO$_2$
film on a sapphire wall at 5~$\mu$m separation are shown by the
solid and dashed lines as the functions of wall temperature $T_W$
in thermal equilibrium (the environment temperature $T_E = T_W$,
the upper pair of solid and dashed lines computed with disregarded
and taken into account dc conductivity, respectively, and the top
solid line for the metallic phase) and out of thermal equilibrium
($T_E = 300~$K, the lower pair of solid and dashed lines computed
with disregarded and taken into account dc conductivity,
respectively, and the bottom solid line for the metallic phase).
\label{fig3}}
\end{figure}
Now, using the same equations as explained above, we compute the nonequilibrium
Casimir-Polder force between a Rb atom and a VO$_2$ film on a sapphire wall
spaced at $a=5~\mu$m separation as a function of wall temperature.
The computational results are shown in Figure~\ref{fig3} by the lower pair of
solid and dashed lines (wall temperature varies from 300~K to 340~K, VO$_2$
is in the dielectric phase, the dc conductivity is disregarded and taken into
account, respectively) and by the bottom solid line
(wall temperature varies from 342~K to 380~K, VO$_2$
is in the metallic phase). This computation is made at the environment temperature
$T_E=300~$K. For comparison purposes, the upper pair of solid and dashed lines
and the top solid line in the region of higher temperatures show similar
equilibrium results computed under a condition $T_E=T_W$, i.e., when the
environment is heated up to the same temperature as the wall (measurement
of the difference Casimir force in such a situation was proposed in \cite{82}).

As is seen in Figure~\ref{fig3}, the jump in the Casimir-Polder force due to
the phase transition is much more pronounced in the nonequilibrium case than
in the equilibrium one. Thus, the equilibrium Casimir-Polder force at room
temperature ($T_E=T_W=300~$K), just before the phase transition ($T_E=T_W=340~$K)
and just after the phase transition ($T_E=T_W=342~$K) are equal to --1.92,
--2.17, and --2.68~($10^{-13}$fN), respectively (the dc conductivity in the dielectric
phase is disregarded). If, however, the situation can be nonequilibrium, i.e.,
$T_E=300~$K, but $T_W=300~$K, 340~K, and 342~K, one finds larger in magnitude
values of the Casimir-Polder force  --1.92,
--5.05, and --8.29~($10^{-13}$fN), respectively (the dc conductivity is again
 disregarded).

One more conclusion following from  Figure~\ref{fig3} does not depend on whether the
plate material undergoes the phase transition or is an ordinary dielectric or metal. In
all these cases the magnitude of the equilibrium Casimir-Polder force obtained when both
the plate and the environment are heated up to some temperature $T$ > 300~K is
substantially smaller than the magnitude of the nonequilibrium force in the case
when only the plate is heated up to the temperature $T$ whereas the environment
preserves its temperature of 300~K.

\begin{figure}[!h]
\centering
\vspace*{-2.5cm}
\includegraphics[width=15 cm]{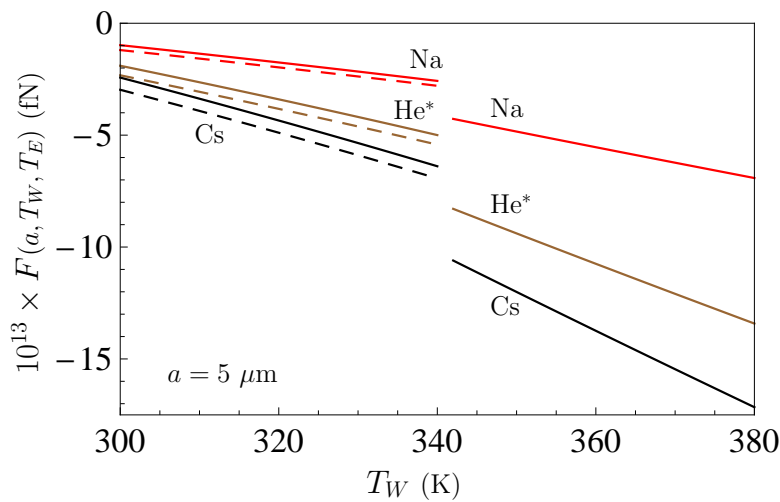}
\vspace*{-12cm}
\caption{The nonequilibrium Casimir-Polder forces between Na,
He$^*$, and Cs atoms and VO$_2$ film on a sapphire wall at
5~$\mu$m separation are shown by the upper, middle, and lower
pairs of solid and dashed lines, respectively, and respective
solid lines as the functions of wall temperature at the environment
temperature $T_E = 300~$K. The solid and dashed lines incorporated
in pairs are computed with disregarded and taken into account dc
conductivity, respectively.
\label{fig4}}
\end{figure}

The nonequilibrium Casimir-Polder force as a function of wall temperature was
also computed for atoms of Na, He$^{\ast}$ and Cs spaced 5~$\mu$m apart of
VO$_2$ film on a sapphire wall, whereas the environment was kept at  $T_E=300~$K.
The results are presented in Figure~\ref{fig4} in the same way as for Rb atoms
in the nonequilibrium case in Figure~\ref{fig3}. The three pairs of solid and
dashed lines from top to bottom are for Na, He$^{\ast}$ and Cs atoms
interacting with a VO$_2$ film in the dielectric phase (as above, the solid lines
disregard dc conductivity and the dashed lines take it into account).
The three solid lines from top to bottom are for the same respective atoms
interacting with a VO$_2$ film in the metallic phase. In all these cases the
magnitude of the Casimir-Polder force increases significantly in
out-of-thermal equilibrium conditions.

\section{ Comparison between the Nonequilibrium Casimir-Polder Forces for
the Phase-Change and Dielectric Materials}

{}From Figures \ref{fig1}, \ref{fig3}, and \ref{fig4}, it is seen that both the equilibrium
and nonequilibrium Casimir-Polder forces are extremely small. In fact the forces
below 1~fN are not accessible to a direct experimental observation.
However, as noted in Section~1, the atom-wall forces lead to measurable
effects in precise experiments with atomic clouds incorporating a great number
of atoms. Thus, the Bose-Einstein condensate of $^{87}$Rb atoms was produced
in a magnetic trap near a silica glass (SiO$_2$) surface and resonantly
driven into a dipole oscillation \cite{49}. It was shown \cite{47,48} that
both in the equilibrium ($T_W=T_E$) and nonequilibrium ($T_W>T_E$) situations
the Casimir-Polder forces lead to quite measurable shifts in the oscillation
resonant frequency. These shifts were measured and recalculated into the
gradients of the Casimir-Polder force \cite{49}.

Below we compare the values of the nonequilibrium Casimir-Polder forces and
their gradients, which were obtained in the experimental configuration \cite{49}
using the dielectric SiO$_2$ wall, with those found for the sapphire wall
covered with a VO$_2$ film which undergoes the dielectric-to-metal phase
transition with increasing temperature. The computations of the Casimir-Polder
force in both cases are made using Equations (\ref{eq1})--(\ref{eq7}) where
the values of all necessary parameters are indicated above. For the static
dielectric permittivity of  SiO$_2$ one has
$\varepsilon^{\rm SiO_2}(0)=3.8$ \cite{83}.
The temperature of the environment $T_E=310~$K is used as in \cite{49}.

\begin{figure}[!h]
\centering
\vspace*{-2.5cm}
\includegraphics[width=15 cm]{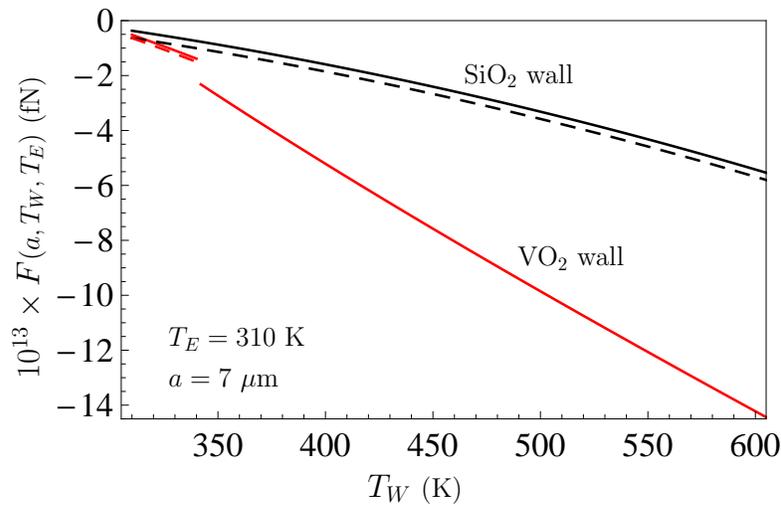}
\vspace*{-12cm}
\caption{The nonequilibrium Casimir-Polder forces between Rb
atoms and a wall at 7~$\mu$m separation are shown as the functions
of wall temperature by the upper pair of solid and dashed lines
for a SiO$_2$ wall and by the lower pair of solid and dashed lines
continued to higher temperatures by the bottom solid line for a
VO$_2$ film on a sapphire wall (the environment temperature is
$T_E = 310~$K). The solid and dashed lines incorporated in pairs
are computed with disregarded and taken into account dc
conductivity of dielectric materials.
\label{fig5}}
\end{figure}

In Figure~\ref{fig5}, the nonequilibrium Casimir-Polder forces between a Rb
atom and a wall at 7~$\mu$m separation are shown as the functions
of wall temperature by the upper pair of solid and dashed lines
for a SiO$_2$ wall and by the lower pair of solid and dashed lines
continued to higher temperatures by the bottom solid line for a
VO$_2$ film deposited on a sapphire wall. In both cases the solid and dashed lines
incorporated in pairs are computed with disregarded and taken into account dc
conductivity of a dielectric material, respectively.
As is seen in Figure~\ref{fig5}, in the case of phase-change wall material the
magnitudes of the Casimir-Polder force reach much larger values than for
a dielectric SiO$_2$ wall. For example, at $T_W=415~$K the Casimir-Polder force
between a Rb atom and a VO$_2$ film on a sapphire wall reaches the same magnitude
as it reaches at $T_W=605~$K in the case of a SiO$_2$ wall.

\begin{figure}[!t]
\centering
\vspace*{-2.5cm}
\includegraphics[width=15 cm]{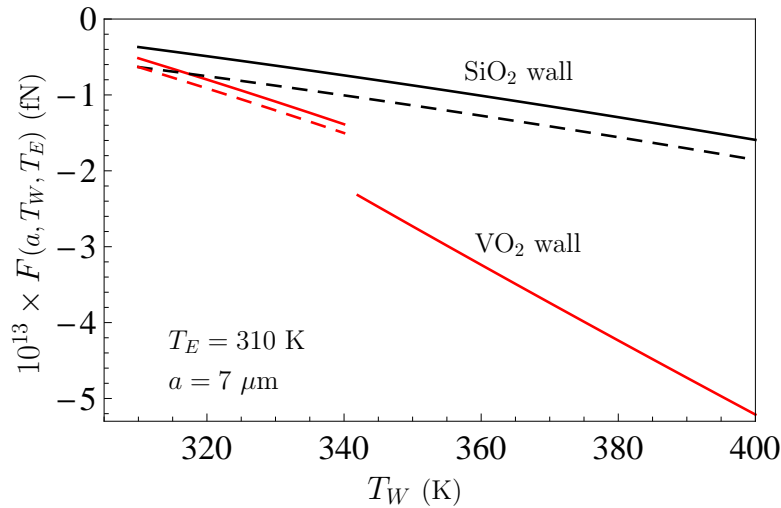}
\vspace*{-12cm}
\caption{The nonequilibrium Casimir-Polder forces between Rb
atoms at 7~$\mu$m separation from the wall made either of SiO$_2$
or sapphire coated with VO$_2$ film are shown as the functions
of wall temperature in the vicinity of the critical temperature.
All notations are the same as in the caption of Figure~\ref{fig5}.
\label{fig6}}
\end{figure}
In Figure~\ref{fig6}, the comparison between the Casimir-Polder forces in the
cases of phase-change and dielectric walls is made on an enlarged scale in the
vicinity of the transition temperature $T_c=341~$K. It is seen that although
the phase transition contributes essentially to the magnitude of the
Casimir-Polder force at $T>T_c$, the slope of the force lines for a phase-change
wall is in any case larger than for a dielectric wall. This makes the phase-change
wall material preferable for measurements of the nonequilibrium Casimir-Polder
force in precise experiments.

To confirm this conclusion, we also compare the gradients of the Casimir-Polder forces
between Rb atoms and either the VO$_2$ on sapphire or SiO$_2$ walls taking into
account that just the force gradients are directly connected with the measured frequency
shift \cite{49}. The gradient of the nonequilibrium Casimir-Polder force is obtained
by differentiating (\ref{eq1}) with respect to separation
\begin{equation}
F^{\prime}(a,T_W,T_E)=F_{\rm eq}^{\prime}(a,T_E)+F_{\rm neq}^{\prime}(a,T_W)
-F_{\rm neq}^{\prime}(a,T_E).
\label{eq9}
\end{equation}

Here, the equilibrium contribution to the force gradient for a dielectric wall with
disregarded dc conductivity is obtained from (\ref{eq2})
\begin{equation}
F_{\rm eq}^{\prime}(a,T)=\frac{3k_B T}{a^5}\,\alpha(0)\,
\frac{\varepsilon(0)-1}{\varepsilon(0)+1}
\label{eq10}
\end{equation}
\noindent
and with taken into account  dc conductivity --- from (\ref{eq3})
\begin{equation}
F_{\rm eq}^{\prime}(a,T)=\frac{3k_B T}{a^5}\,\alpha(0).
\label{eq11}
\end{equation}
\noindent
The last result is also valid for a wall material in the metallic phase.

The nonequilibrium contribution to (\ref{eq9}) is obtained
from (\ref{eq4}) for a dielectric wall
\begin{equation}
F_{\rm neq}^{\prime}(a,T)=\frac{\pi\alpha(0)(k_B T)^2}{2c\hbar a^4}\,
\frac{\varepsilon(0)+1}{\sqrt{\varepsilon(0)-1}}
\label{eq12}
\end{equation}
\noindent
and from (\ref{eq6}) for a metallic wall
\begin{equation}
F_{\rm neq}^{\prime}(a,T)=\frac{3\alpha(0)\zeta(3/2)\sqrt{\sigma_m(T)\,}
(k_B T)^{3/2}}{c\sqrt{2\hbar}\, a^4}.
\label{eq13}
\end{equation}

\begin{figure}[!t]
\centering
\vspace*{-2.5cm}
\includegraphics[width=15 cm]{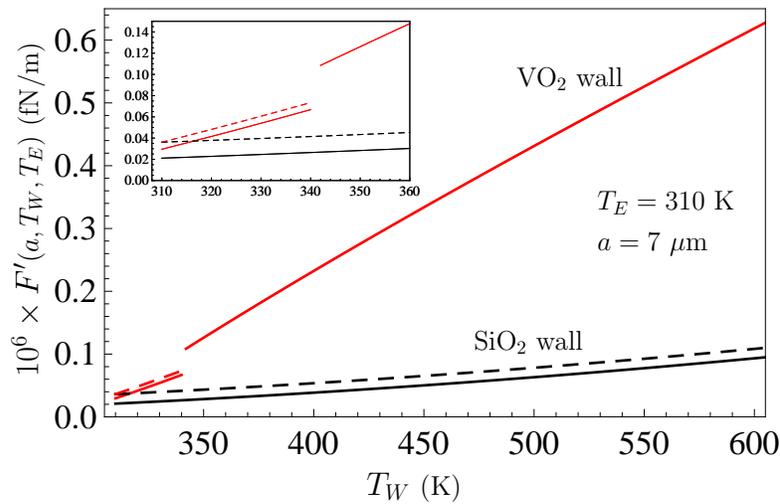}
\vspace*{-12cm}
\caption{The gradients of nonequilibrium Casimir-Polder forces
between Rb atoms and a wall at 7~$\mu$m separation are shown as
the functions of wall temperature by the lower pair of solid and
dashed lines for a SiO$_2$ wall and by the upper pair of solid and
dashed lines continued to higher temperatures by the top solid line
for a VO$_2$ film on a sapphire wall (the environment temperature
is $T_E = 310~$K). The solid and dashed lines incorporated in pairs
are computed with disregarded and taken into account dc
conductivity of dielectric materials. The temperature region in
the vicinity of the critical temperature is shown in the inset on
an enlarged scale.
\label{fig7}}
\end{figure}

Computations of the gradient of the nonequilibrium Casimir-Polder force
between a Rb atom and either a SiO$_2$ wall or a VO$_2$ film on a sapphire
wall at 7~$\mu$m separation were made by Equations (\ref{eq9})--(\ref{eq13}).
The computational results are shown in  Figure~\ref{fig7} as the functions
of wall temperature by the lower pair of solid and dashed lines
for a SiO$_2$ wall and by the upper pair of solid and dashed lines
continued to higher temperatures by the top solid line for a
VO$_2$ film deposited on a sapphire wall. In both cases the solid and dashed lines
incorporated in pairs are computed with disregarded and taken into account dc
conductivity of the dielectric material, respectively.
The environment temperature $T_E=310~$K is used in computations.
In the inset, the range of temperatures in the vicinity of the critical
temperature $T_c=341~$K, at which the phase transition occurs, is shown on
an enlarged scale.

As is seen in  Figure~\ref{fig7}, in the case of a VO$_2$ film on a sapphire wall
one obtains much larger gradients of the nonequilibrium Casimir-Polder force than
for a SiO$_2$ wall under the same conditions. Thus, at $T_W=350~$K and 400~K
the gradients of the  Casimir-Polder force for  a VO$_2$ film on a sapphire wall
exceed those for a SiO$_2$ wall by the factors of 4.46 and 6.02, respectively.
The value of the force gradient reached in the experiment \cite{49} at
$T_W=605~$K would be reached for a VO$_2$ film on a sapphire wall at much
smaller temperature $T_W=342~$K just after the phase transition (compare with
415~K found for the respective values of the magnitude of the Casimir-Polder force
in  Figure~\ref{fig5}). This confirms that the use of phase-change wall materials
is advantageous in experiments on measuring the nonequilibrium Casimir-Polder
forces.

\section{Discussion}

In the foregoing, we have considered the Casimir-Polder interaction
between different atoms and a wall in out-of-thermal-equilibrium
situation when the wall temperature differs from the temperature
of the environment. The Lifshitz theory of the fluctuation-induced
forces has already been generalized for this case \cite{47,48}, and
the nonequilibrium Casimir-Polder force was measured in the pioneer
experiment \cite{49}. The main novel feature of our study is a
suggestion to use the material of the wall which undergoes the
dielectric-to-metal phase transition with increasing wall temperature.

We have considered He$^*$, Na, Rb, and Cs atoms interacting with a
VO$_2$ film deposited on a sapphire wall. The question arises what
is an advantage of vanadium dioxide as compared with silica glass
(SiO$_2$) used in the already performed measurements of the
nonequilibrium Casimir-Polder force. The point is that at relatively
low temperature $T$ = 341K vanadium dioxide undergoes a transition
from the dielectric to metallic phase accompanied by a jump in its
conductivity by the four orders of magnitude. As a result, the
Casimir-Polder force is subjected to a combined action of the
nonequilibrium conditions and the phase transformation. Our
calculations of the nonequilibrium Casimir-Polder force as a
function of atom-wall separation and wall temperature show that in the
case of a phase change wall material the force magnitude reaches
much larger values at the same separation and temperature than in
the case of a dielectric wall.

A comparison of this kind was made not only for the Casimir-Polder
force but for its gradient as well taking into account that just
the gradient is connected with the immediately measured shift in
the oscillation resonant frequency of the condensate atomic cloud
in \cite{49}. It is shown that the combined action of the thermal
nonequilibrium and phase transition on the force gradient is even
more pronounced than on the Casimir-Polder force. As an example,
at 7~$\mu$m atom-wall separation and 400~K wall temperature the
force gradient for the phase change wall material is by a factor of
6 greater than for a dielectric wall (whereas the force magnitude
in the same case is greater only by a factor of 3). This opens up
new opportunities for experimental investigation of atom-wall
interaction in out of thermal equilibrium conditions.
\section{Conclusions}

To conclude, in spite of extremely small magnitudes of the Casimir-
Polder forces acting between separate atoms and cavity walls, they
lead to important effects which are observable in such physical
phenomena as quantum reflection and Bose-Einstein condensation
dealing with clouds embodying the great number of cold atoms.
The use of out-of-thermal-equilibrium conditions enhances the
capabilities for investigation of atom-wall interaction. We have shown
that heating of the wall alone by keeping the environment temperature
constant leads to larger magnitudes of the nonequilibrium Casimir-Polder
force, as compared to the equilibtium ones obtained when both the wall
and the environment are heated up to the same temperature. According
to our results, even more prospective possibilities for research
in the area of Casimir-Polder forces are presented by the combined
use of nonequilibrium conditions and phase-change wall materials.

\funding{This work was supported by the Peter the Great Saint Petersburg Polytechnic
University in the framework of the Russian state assignment for basic research
(project N FSEG-2020-0024).
V.M.M.~was partially funded by the Russian Foundation for Basic Research grant number
19-02-00453 A.}

\acknowledgments{
V.M.M.\ was partially supported by the Russian Government Program of Competitive Growth of Kazan Federal University.}


\reftitle{References}

\end{document}